\documentclass[prd,twocolumn,reprint,preprintnumbers,nofootinbib,superscriptaddress]{revtex4-2}
\input{header.tex}

\begin{document}
\reportnum{-2}{CERN-TH-2025-158}
\title{Precision calculation of $N_{\text{eff}}$ with Neutrino Direct Simulation Monte Carlo}

\author{Oleksii~Ihnatenko}
\email{oleksii.ihnatenko@knu.ua}
\affiliation{Taras Shevchenko National University of Kyiv, 64 Volodymyrs’ka str., Kyiv 01601, Ukraine}
\author{Maksym~Ovchynnikov}
\email{maksym.ovchynnikov@cern.ch}
\affiliation{Theoretical Physics Department, CERN, 1211 Geneva 23, Switzerland}

\date{\today}

\begin{abstract}
Neutrino Direct Simulation Monte Carlo ($\nu$DSMC) is a Monte Carlo method for solving the neutrino Boltzmann equation in the early Universe, designed to track the evolution of cosmic neutrinos across a wide range of cosmological scenarios. We develop a complete $\nu$DSMC solver that consistently incorporates the effects of the electron mass, three-flavour neutrino oscillations, and finite-temperature QED corrections to the thermodynamics of the electromagnetic plasma. As a first application, we perform a high-precision calculation of neutrino decoupling in the standard cosmological model and obtain
$N_{\text{eff}} = 3.0439 \pm 0.0006$, in excellent agreement with state-of-the-art results.
\end{abstract}

\maketitle

\section{Introduction}
\label{sec:introduction}

Cosmic neutrinos ($\nu$s) are a crucial component of the Early Universe, significantly influencing its evolution during the radiation-dominated epoch. In particular, they constituted a large fraction of the energy density of the thermal plasma at plasma temperatures in the range eV-MeV, and participated in weak processes governing the interconversion of protons and neutrons. This way, they affected the primordial nuclear abundances and formation of the Cosmic Microwave Background,~\cite{Dolgov:2002wy,Lesgourgues:2006nd,ParticleDataGroup:2024cfk}. Altogether, understanding the dynamics of neutrinos is essential for revealing the cosmological history.

According to the standard cosmological model ($\Lambda$CDM), $\nu$s were in thermal equilibrium at temperatures $T\gg \SI{1}{\MeV}$, then gradually underwent decoupling from the electromagnetic (EM) plasma at around \si{\MeV} temperatures and became cosmic relics afterward. One of the simplest quantities characterizing this dynamics is the effective number of ultrarelativistic degrees of freedom, or $N_{\text{eff}}$:
\begin{equation}
    N_{\text{eff}} = \frac{8}{7}\left(\frac{11}{4}\right)^{\frac{4}{3}}\frac{\rho_{\text{radiation}}-\rho_{\gamma}}{\rho_{\gamma}}\bigg|_{T = T_{\text{rec}}},
    \label{eq:Neff}
\end{equation}
where $T_{\text{rec}}$ is the recombination temperature, while $\rho_{\text{radiation},\gamma}$ are energy densities of all ultrarelativistic species (including photons and neutrinos) and photons, respectively. In $\Lambda$CDM, the numerator matches the neutrino energy density $\rho_{\nu}$, which is a sum over neutrinos $\nu_{\alpha}$ and antineutrinos $\bar{\nu}_{\alpha}$ of different flavors $\alpha = e,\mu,\tau$. 

The status of $N_{\text{eff}}$ measurements is as follows. Observations by Planck~\cite{Planck:2018vyg}, assuming the $\Lambda$CDM model, provide $N_{\text{eff}} = 2.99\pm 0.17$ at $1\sigma$ CL; if additionally marginalizing over the helium abundance, the measurements provide $N_{\text{eff}} = 2.89\pm 0.31$~\cite{Boyarsky:2021yoh}. Recent observations from ACT~\cite{ACT:2025tim,ACT:2025fju} and SPT-3G~\cite{SPT-3G:2025bzu} leave the $N_{\text{eff}}$ uncertainties comparable. Ongoing observations by Simons Observatory~\cite{SimonsObservatory:2018koc} aim to deliver precision $\Delta N_{\text{eff}} = \pm 0.05$, while the next generation of CMB observations, such as CMB-S4, may reach the precision of 0.014~\cite{CMB-S4:2016ple}.

These measurements should be compared to theoretical calculations of $N_{\text{eff}}$. Under certain approximations about neutrino oscillations, such as their averaging over oscillation length, the task reduces to solving the Boltzmann equation on the distribution function $f_{\nu_{\alpha}}(p,t)$, where $p$ is the modulus of momentum, and $t$ is time. Explicitly,
\begin{equation}
    \partial_{t}f_{\nu_{\alpha}} - Hp\partial_{p}f_{\nu_{\alpha}} = \mathcal{I}_{\text{coll},\nu_{\alpha}}[f_{\nu_{\alpha}},p],
    \label{eq:boltzmann}
\end{equation}
with $H$ being the Hubble factor, and $\mathcal{I}_{\text{coll},\nu_{\alpha}}$ is the collision integral. 

Precision calculations of $N_{\text{eff}}$ within $\Lambda$CDM have been performed by two approaches. They are the integrated Boltzmann equation method (where the shape of the neutrino distribution is approximated by the Fermi-Dirac one)~\cite{Escudero:2018mvt,EscuderoAbenza:2020cmq,Cielo:2023bqp}, and the approach of solving the neutrino Boltzmann equation or quantum kinetic equations via discretization of comoving momentum space~\cite{Hannestad:1995rs,Mangano:2005cc,deSalas:2016ztq,Grohs:2015tfy,Gariazzo:2019gyi,Froustey:2019owm,Bennett:2019ewm,Bennett:2020zkv,Froustey:2021azz,Froustey:2020mcq,Froustey:2024mgf,Drewes:2024wbw}. If including QED corrections to the thermodynamics of the EM plasma, and neutrino oscillations, both methods coherently give $N_{\text{eff}} = \numrange{3.043}{3.044}$, in agreement with the observations.

Large error bars in $N_{\text{eff}}$ measurements, however, leave room for non-standard physics that may have existed during neutrino decoupling. Many different non-standard cosmological scenarios that may lead to a modification of $N_{\text{eff}}$ and the shape of the neutrino distribution function. One example is a non-thermal injection of various Standard Model particles, usually absent at the MeV plasma: extra EM plasma, non-thermal neutrinos, and metastable species such as $\mu,\pi^{\pm},K$. It may happen, e.g., because of decaying thermal relics~\cite{Pospelov:2010cw,Fradette:2018hhl,Gelmini:2020ekg,Sabti:2020yrt,Boyarsky:2020dzc,Boyarsky:2021yoh,Rasmussen:2021kbf,Akita:2024nam,Akita:2024ork} or in late reheating scenarios~\cite{Kawasaki:2000en,Hannestad:2004px,Ichikawa:2005vw,Hasegawa:2019jsa,Barbieri:2025moq}, evaporating primordial black holes~\cite{Acharya:2020jbv,Keith:2020jww}, decaying topological defects, and other scenarios. 

To interpret the current and, especially, upcoming BBN and CMB observations in light of these models, it is necessary to accurately solve the equation~\eqref{eq:boltzmann} beyond $\Lambda$CDM. However, utilizing the state-of-the-art approaches for this purpose may be non-trivial. These events would induce distortions in the neutrino distribution. These distortions are important both quantitatively and qualitatively as they influence the sign of the $N_{\text{eff}}$ change~\cite{Boyarsky:2021yoh,Ovchynnikov:2024rfu,Akita:2024nam}. By definition, they cannot be studied with the integrated approach. On the other hand, one has to simultaneously carefully trace the thermal part of the neutrino spectrum (fixed in comoving momentum space) and the non-thermal products (for which the comoving spectrum heavily evolves, depending on the temperature). To achieve this, the discretization approach requires very fine binning in momentum space, which quickly makes it unfeasible if the injected neutrinos are very energetic due to the growth of the computational time~\cite{Sabti:2020yrt,Akita:2024ork}. The situation becomes especially complicated if jet decay products are present: they undergo hadronization. The phase space of the resulting products is complicated to describe analytically and is not easily fit within the collision-integral-reduction schemes of the momentum discretization approaches.

In Refs.~\cite{Ovchynnikov:2024rfu,Ovchynnikov:2024xyd}, a novel method to study the neutrino decoupling based on the so-called Direct Simulation Monte Carlo (DSMC) has been proposed -- $\nu$DSMC. Instead of solving the Boltzmann equation~\eqref{eq:boltzmann} explicitly, $\nu$DSMC starts with a huge number $N\gg 1$ of neutrinos, EM particles, and non-standard species, and simulates their interactions throughout the expansion of the Universe using Monte-Carlo techniques. This way, it naturally avoids the need to discretize momentum and analytically present the phase space of hadronic decays. Additionally, energy is conserved exactly in every individual collision. On the other hand, the DSMC interaction kernel delivers good performance even for a huge number of particles $N\sim 10^{7}$.

A prototype of $\nu$DSMC, developed in~\cite{Ovchynnikov:2024rfu,Ovchynnikov:2024xyd}, lacked several important features required to apply it in full generality, notably the finite electron mass and QED corrections to the thermodynamics of the electromagnetic plasma. The latter are essential for making accurate calculations of $N_{\text{eff}}$ independently of the cosmological setup.

In this study, we develop the full version of $\nu$DSMC by incorporating the electron mass, neutrino oscillations, and QED corrections up to the order of $\mathcal{O}(e^{3})$, and then apply it to precision calculations of neutrino decoupling in $\Lambda$CDM.\footnote{The $\nu$DSMC code will be published in a dedicated study, but may be provided upon request.} Aside from improving the maturity of $\nu$DSMC and making it ready for various non-standard applications, our results serve as an independent cross-check of the various state-of-the-art approaches under different approximations of the standard cosmological setup.

The paper is organized as follows. Sec.~\ref{sec:vanilla-DSMC} introduces the traditional DSMC approach used in the field of physics of rarefied gases. Sec.~\ref{sec:nuDSMC} discusses the conceptual modifications in DSMC required for describing the interactions of cosmic neutrinos in the expanding Universe. Sec.~\ref{sec:results} discusses cross-checks of our approach, its application to the standard cosmological scenario of neutrino decoupling, and its role in the context of current and future observations. Finally, we conclude in Sec.~\ref{sec:conclusions}.

\section{Traditional DSMC}
\label{sec:vanilla-DSMC}
Let $F_{N}(\mathcal{R},\mathcal{V},t)$ be the $N$-particle probability density on phase space, where
$\mathcal{R}=(\mathbf{r}_1,\ldots,\mathbf{r}_N)$ and $\mathcal{V}=(\mathbf{v}_1,\ldots,\mathbf{v}_N)$ are spatial coordinates and velocities. Assume short‑range pair interactions generated by a symmetric potential $\phi_{ij}=\phi(|\mathbf{r}_i-\mathbf{r}_j|)$. The dynamics of $F_{N}(\mathcal{R},\mathcal{V},t)$ obeys the Liouville equation
\begin{equation}
  \frac{\partial F_N}{\partial t}
  + \sum_{i=1}^N \mathbf{v}_i \cdot \nabla_{\mathbf{r}_i} F_N
  + \sum_{i=1}^N \frac{1}{m_i}\,\mathbf{F}_i(\mathcal{R}) \cdot \nabla_{\mathbf{v}_i} F_N
  = 0,
  \label{eq:liouville}
\end{equation}
with $\mathbf{F}_i(\mathcal{R}) \equiv - \sum_{\substack{j=1\\ j\neq i}}^N \nabla_{\mathbf{r}_i}\phi_{ij}$, subject to suitable initial and boundary conditions.

Next, let us define the reduced $k$‑particle distributions
\begin{equation}
  f^{(k)}(\mathbf{r}_1,\mathbf{v}_1,\ldots,\mathbf{r}_k,\mathbf{v}_k,t)
  = \int \prod_{j = k+1}^{N}\mathrm{d}\mathbf{r}_{j} \mathrm{d}\mathbf{v}_{j} F_N
  \label{eq:marginals}
\end{equation}
Under standard assumptions, these marginals satisfy a Bogoliubov-Born-Green-Kirkwood-Yvon-type hierarchy. In the Boltzmann-Grad limit, together with the hypothesis of the statistical independence of pre-collisional velocities within local subdomains (local molecular chaos), the hierarchy closes at $k=1$ and reduces to the Boltzmann equation~\cite{bird1970direct,ivanov1991theoretical,stefanov2019basic}.

The DSMC approach approximates solutions of~\eqref{eq:liouville} on time intervals $(t,t+\Delta t)$. The method proceeds through the following steps:
\begin{itemize}
  \item[--] \textit{Time discretization.} Choose $\Delta t$ small compared with the mean collision time. Over $\Delta t$, the evolution is split into a transport sub-step (where external forces are applied to the particles) and a collision sub-step (Lie-Trotter or Strang composition yields first‑ or second‑order accuracy, respectively).
  \item[--] \textit{Spatial discretization.} Partition the physical domain $\mathcal{D}$ of volume $V$ into $M$ disjoint cells $\{\mathcal{D}^{(l)}\}_{l=1}^{M}$ with volumes $V_{\text{cell}}^{(l)}$. During the collision sub-step, the set of simulation particles resident in a cell $\mathcal{D}^{(l)}$ is frozen; collisions are assumed spatially homogeneous inside the cell.
  \item[--] \textit{Collision modelling in a cell.} With $N_{\text{cell}}$ simulation particles present, unordered pairs are drawn uniformly from the $\binom{N_{\text{cell}}}{2}$ possibilities. Each pair is tested for interaction by an accept-reject rule against a cell‑wise majorant of the collision frequency (the No-Time-Counter/Majorant-Collision-Frequency kernel).
\end{itemize}

Pairs are accepted with probability
\begin{equation}
  P_{\text{acc}} = \frac{(\sigma v)_{\text{pair}}}{(\sigma v)_{\text{cell,max}}},
  \label{eq:vanilla-DSMC-acceptance}
\end{equation}
where $(\sigma v)_{\text{pair}}$ is the product of the pair cross‑section and their relative velocity, and $(\sigma v)_{\text{cell,max}}$ is a uniform majorant within the cell.
For relativistic $2\!\to\!2$ collisions, the invariant cross‑section is~\cite{Weinberg:1995mt}
\begin{equation}
  \sigma_{12\to 34}
  = \int \mathrm{d}\cos\theta_{\text{CM}}\,
      \frac{\pi |\mathbf{p}_{3}|\,
            \overline{\left|\mathcal{M}_{12\to 34}\right|^{2}}}
           {8E_{1}E_{2}\sqrt{s}\,v},
  \label{eq:cross-section-inv}
\end{equation}
and the Lorentz‑invariant relative velocity reads
\begin{equation}
  v = \frac{\sqrt{(p_{1}\!\cdot\!p_{2})^{2}-m_{1}^{2}m_{2}^{2}}}{E_{1}E_{2}}.
  \label{eq:velocity}
\end{equation}
Using Eqs.~\eqref{eq:cross-section-inv}-\eqref{eq:velocity} guarantees convergence to the correct relativistic
equilibrium~\cite{peano2009statistical}.

For the No-Time-Counter kernel~\cite{bird1989perception} (which we will utilize below), the \emph{expected} number of candidate pairs tested in a cell per timestep is
\begin{multline}
  \mathbb{E}[N_{\text{pairs}}]
  = \binom{N_{\text{cell}}}{2}\,
    \frac{(\sigma v)_{\text{cell,max}}}{V_{\text{cell}}}\,\Delta t
  = \\ = \frac{N_{\text{cell}}\!\left(N_{\text{cell}}-1\right)}{2}\,
    \frac{(\sigma v)_{\text{cell,max}}}{V_{\text{cell}}}\,\Delta t,
  \label{eq:ntc-budget}
\end{multline}
with $V_{\text{cell}}$ the cell volume. In practice, one draws $N_{\text{pairs}}$ as a non‑negative integer with mean~\eqref{eq:ntc-budget} (e.g., a Poisson variate), then applies the acceptance test~\eqref{eq:vanilla-DSMC-acceptance}. Upon acceptance, post‑collisional kinematics and, where appropriate, particle types and multiplicities for $2\!\to\!n$ scattering processes are sampled from the differential cross‑section. Inter‑cell effects are handled solely by the subsequent transport sub-step.

\section{$\nu$DSMC}
\label{sec:nuDSMC}

In this section, we formulate the complete $\nu$DSMC approach, discussing the dynamics of the Universe expansion, QED corrections, interactions of neutrinos and EM particles, and neutrino oscillations.

\subsection{Expanding Universe and $\nu$DSMC input}
\label{sec:universe}
Let us start by discussing the global properties of the expanding Universe at MeV temperatures, and how they change the traditional DSMC routine.

The Universe is highly homogeneous and isotropic, which means that we do not need to store particles' coordinates, and their relative velocities are isotropically distributed. This leads to a number of simplifications. First, effectively, we may simulate the interactions as if all the particles were meeting at one point. Second, there is no need to track the individual particles' 4-momenta: it is enough to characterize their kinematics solely by energies, and sample directions randomly (we will return to this in Sec.~\ref{sec:interaction-kernel}). Finally, the boundary cell-cell interactions during the interaction stage may be completely neglected, and after the interaction, the distribution of particles between the cells is simply randomly shuffled. 

In $\nu$DSMC, the expansion itself plays the role of an external force: at the transport sub-step, we reduce particles' momenta, and on top of that, inflate the volume of the system (the cosmological redshift). 

Now, let us clarify the dynamics of the Universe's expansion. In the standard cosmological scenarios, it is dominated by the evolution of neutrinos and EM particles. Dropping the $\nu$-EM interactions and treating the plasma components as a perfect fluid, we may write the following continuity equations on their energy densities $\rho$:
\begin{align}
    &\frac{d\rho_{\nu}}{dt} + 4H\rho_{\nu} = 0, \\ &\frac{d\rho_{\text{EM}}}{dt} + 3H(\rho_{\text{EM}}+P_{\text{EM}}) = 0
    \label{eq:redshift-continuous}
\end{align}
Here, $P$ is pressure, ``$\nu$'' sums up over three active flavors, while ``$\text{EM}$'' over electrons, positrons, and photons. $H$ is the Hubble factor:
\begin{equation}
    H = \sqrt{\frac{8\pi}{3m_{\text{Pl}}^{2}}(\rho_{\nu}+\rho_{\text{EM}})},
    \label{eq:Hubble}
\end{equation}
with $m_{\text{Pl}} \approx 1.2\cdot 10^{19}\text{ GeV}$ being Planck's mass. It defines the expansion law of the volume: $dV/dt -3HV = 0$. The expressions for the EM energy density and pressure are discussed in detail in Sec.~\ref{sec:thermal-corrections}.

Consequently, in $\nu$DSMC, incorporation of the cosmological expansion is required. On top of the requirement to resolve the interactions, the timestep $\Delta t$ now has to be chosen such that $\Delta t \ll H^{-1}$. At the transport sub-step, we redshift the neutrino energies $E_{\nu}$, the system's volume $V$, and the EM plasma energy density $\rho_{\text{EM}}$ using the discrete versions of the laws~\eqref{eq:redshift-continuous}:
\begin{align}
E_{\nu}&\to E_{\nu}\exp[-H\Delta t], \quad V \to V\exp[3H\Delta t], \\ \rho_{\text{EM}}&\to \rho_{\text{EM}}\exp\left[-3H\Delta t\frac{\rho_{\text{EM}}+P_{\text{EM}}}{\rho_{\text{EM}}}\right]
\label{eq:expansion-discretized}
\end{align}
In thermal equilibrium (which is the case at temperatures $T\gg 1\text{ MeV}$), the $\nu$ and EM particles are described by the Fermi-Dirac and Bose-Einstein distributions with the same temperatures $T_{\nu_{\alpha}}=T_{\text{EM}}$ (further, we denote $T_{\text{EM}}\equiv T$). We may collectively characterize them solely by $\rho_{\nu},\rho_{\text{EM}}$ even if including interactions, without any need for the approaches tracking the neutrino energy distribution function. However, the $\nu-\nu$ and $\nu-\text{EM}$ interactions are much less efficient than the EM-EM processes, and at MeV temperatures (where the EM-EM processes are still very fast) they become comparable with the Hubble expansion rate, meaning that neutrinos start decoupling. 

This necessitates introducing a hybrid description of the neutrino and EM particles within the $\nu$DSMC. Namely, neutrinos must be represented by simulation particles characterized by the lepton flavor and energy. As for the EM particles, introducing simulation electrons and positrons makes no sense: their energies and amount would be collectively updated according to the dynamics of the EM plasma temperature. Instead, we characterize the EM plasma solely by the temperature $T$, both globally and at the cell level (with each cell having its own temperature $T_{\text{cell}}$). 

Therefore, the input to $\nu$DSMC is a set of neutrinos in the form of their flavor $\alpha = e,\mu,\tau$ and energy $E$, the initial volume of the system $V$, and the temperature $T$ of the EM plasma. Performing sequential cosmological redshift and interactions (discussed in Sec.~\ref{sec:interaction-kernel}), we obtain as an output the updated set of neutrinos, the evolution $V(t)$ (and hence scale factor of the Universe $a(t)$, as $V \propto a^{3}$) and $T(t)$. In the equilibrium, the shape of the neutrino energy distribution obtained by collecting all neutrinos must match $E^{2}\times f_{\text{FD}}(T_{\nu_{\alpha}},E)$, where $E^{2}$ comes from the phase space measure~\cite{peano2009statistical}, and the neutrino temperatures $T_{\nu_{\alpha}} \equiv T$.

Summing over neutrinos energies, we get $N_{\text{eff}}$:
\begin{equation}
    N_{\text{eff}} = \frac{8}{7}\left(\frac{11}{4}\right)^{\frac{4}{3}}\frac{E_{\nu,\text{tot}}}{E_{\text{EM,tot}}}\bigg|_{T\ll m_{e}},
    \label{eq:Neff-DSMC}
\end{equation}
where $E_{i,\text{tot}}$ is the total energy of the species $i$.

Finally, let us comment on the modifications of the routine due to the presence of hypothetical unstable new physics particles $Y$. They add the term $\rho_{Y}$ to the energy density entering the Hubble factor~\eqref{eq:Hubble}. We sample their decay times throughout the timespan of the evolution using the exponential distribution $dN/dt = -N(t)/\tau$, and at every timestep $(t,t+\Delta t)$, we may generate decays of the corresponding number of particles using either custom decay sampler or state-of-the-art approaches in particle physics, such as \texttt{PYTHIA8}~\cite{Bierlich:2022pfr}. The decays are described by adding neutrinos to the simulation pool (for decays into neutrinos), increasing the EM energy density (for purely EM decays), or both (for decays into metastable particles with temperature-dependent decay probability~\cite{Akita:2024nam,Akita:2024ork}). 

\subsubsection{Thermodynamics of the EM plasma}
\label{sec:thermal-corrections}
Describing the EM plasma in terms of $T$ requires knowing the relation $\rho,P = \rho(T), P(T)$, as well as the relation $T = T(\rho_{\text{EM}})$, which is needed to trace the effects of the interaction with neutrinos on the EM plasma temperature inside each cell. These relations are given by the equilibrium description of photons and electrons by the Bose-Einstein and Fermi-Dirac distributions, plus non-negligible quantum corrections to thermodynamic quantities coming from self-interactions of the EM particles. The total pressure and energy density of the EM plasma are given by
\begin{equation}
    P_{\text{EM}} = P^{0}_{\text{EM}} + \sum_{i}\delta P^{(i)}_{\text{EM}}, \quad \rho_{\text{EM}} = \rho^{0}_{\text{EM}} + \sum_{i}\delta \rho^{(i)}_{\text{EM}}, 
    \label{eq:EM-thermodynamics}
\end{equation}
where $(i)$ means the order in the perturbative $e^{i}$ expansion. The consistency of the expansion implies the following thermodynamic relation between $P$ and $\rho_{\text{EM}}$:
\begin{equation}
    \rho_{\text{EM}} = -P_{\text{EM}} + T\partial_{T}P_{\text{EM}}
    \label{eq:thermodynamic-relation}
\end{equation}
At zeroth order, they are
\begin{equation}
    P_{\text{EM}}^{(0)}= \frac{T}{\pi^{2}}\int \limits_{0}^{\infty}dk \ k^{2}\log\left[ \frac{\left( 1+\exp\left[-\frac{E_{e}}{T}\right]\right)^{2}}{1-\exp\left[-\frac{E_{\gamma}}{T}\right]}\right], 
\end{equation}
with $E_{e} = \sqrt{k^{2}+m_{e}^{2}}$ and $E_{\gamma} = k$. The energy density may then be obtained using the identity~\eqref{eq:thermodynamic-relation}:
\begin{multline}
    \rho_{\text{EM}}^{(0)} = \frac{2}{\pi^{2}}\int dk \ \frac{k^{2}E_{e}}{\exp\left[\frac{E_{e}}{T}\right]+1} + \frac{1}{\pi^{2}}\int dk \ \frac{k^{3}}{\exp\left[\frac{E_{\gamma}}{T}\right]-1}
    \nonumber
\end{multline}
To incorporate the QED corrections up to the order $\mathcal{O}(e^{3})$ in the EM coupling $e$, we mainly follow the approach of Refs.~\cite{Bennett:2019ewm,Akita:2020szl}. Namely, the $\mathcal{O}(e^{2})$ correction to the EM pressure is given by
\begin{multline}
\delta P_{\text{EM}}^{(2)} = -\int \limits_{0}^{\infty} \frac{dk}{2\pi^{2}}\bigg[\frac{k^{2}}{E_{k}}\frac{\delta m_{e(2)}^{2}}{\exp\left[\frac{E_{k}}{T}\right]+1} +\frac{k}{2}\frac{\delta m_{\gamma(2)}^{2}}{\exp\left[\frac{k}{T}\right]-1}\bigg] 
\nonumber
\end{multline}
Here, $E_{k} = \sqrt{k^{2}+m_{e}^{2}}$, and $\delta m^{2}_{e(2)},\delta m_{\gamma(2)}^{2}$ are thermal corrections to the electron and photon masses, which read
\begin{align}
    \delta m^{2}_{e(2)} &= \frac{2\pi\alpha_{\text{EM}}}{3}T^{2} + \frac{4\alpha_{\text{EM}}}{\pi}\int \limits_{0}^{\infty}dk\frac{k^{2}}{E_{k}}\frac{1}{\exp\left[\frac{E_{k}}{T}\right]+1}, \nonumber \\ \delta m^{2}_{\gamma(2)} &= \frac{8\alpha_{\text{EM}}}{\pi}\int \limits_{0}^{\infty}dk\frac{k^{2}}{E_{k}}\frac{1}{\exp\left[\frac{E_{k}}{T}\right]+1}
    \label{eq:thermal-electron-mass}
\end{align}
The corresponding correction to the EM energy density, $\delta \rho_{\text{EM}}^{(2)}$, may be self-consistently obtained from $\delta P^{(2)}$ using the same identity as for $\rho_{\text{EM}}^{(0)}$:
\begin{equation}
    \delta \rho_{\text{EM}}^{(2)} = -\delta P_{\text{EM}}^{(2)} + T\partial_{T}\delta P_{\text{EM}}^{(2)}
\end{equation}
The $\mathcal{O}(e^{3})$ correction only exists for the photon part, and reads
\begin{equation}
    \delta P_{\text{EM}}^{(3)} = \frac{e^{3}T}{12\pi^{4}}(I[T])^{\frac{3}{2}},
\end{equation}
with 
\begin{equation}
    I[T] = 2\int \limits_{0}^{\infty} dk \frac{k^{2}+E_{k}^{2}}{E_{k}}\frac{1}{\exp\left[\frac{E_{k}}{T}\right]+1}
\end{equation}

\subsection{Interaction kernel}
\label{sec:interaction-kernel}
\begin{figure}[t!]
    \centering
    \includegraphics[width=\linewidth]{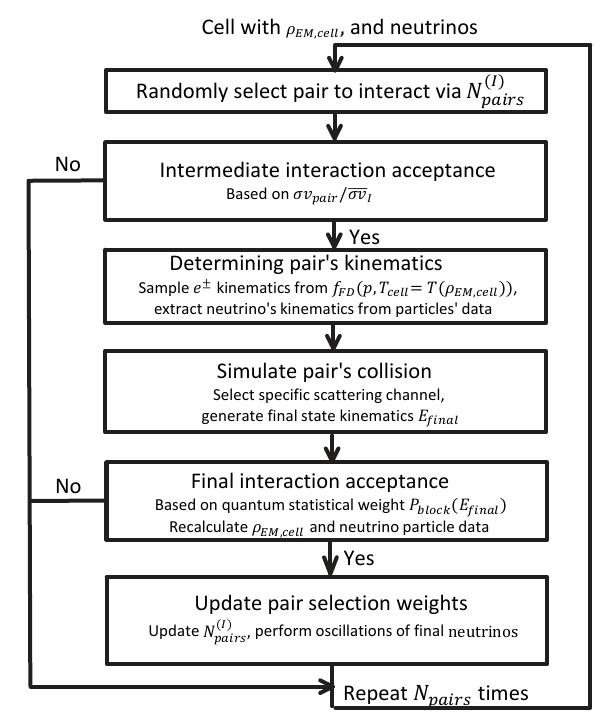}
    \caption{The modification of the No-Time-Counter scheme used in $\nu$DSMC to simulate interactions of neutrinos and EM particles inside cells. First, we perform neutrino oscillations based on the stationary probabilities~\eqref{eq:oscillation-probabilities}. Then, we determine $N_{\text{pairs}}$ pairs of particles to interact -- the sum of Eqs.~\eqref{eq:Npairs-category-1-optimized}-\eqref{eq:Npairs-category-3-optimized}. Then, based on the same equations, iteratively, we decide what type of pair to sample -- $\nu\nu$, $\nu e$, or $ee$. For each pair, we compute its interaction weight $(\sigma v)_{\text{pair}}$ and make an intermediate decision on whether it will interact using the criterion~\eqref{eq:interaction-acceptance-optimized}. Then, we sample the kinematics of the interacting particles, generate the final states resulting from the collision, and make the final decision of whether the interaction takes place from the Pauli principle~\eqref{eq:pauli-blocking}. Finally, we update the local properties of the plasma: the EM plasma temperature and the number of EM particles, as well as neutrino flavor distributions by the oscillation probabilities, Eq.~\eqref{eq:oscillation-probabilities}.}
    \label{fig:new-NTC}
\end{figure}

There are three conceptual features of the NTC kernel in $\nu$DSMC that differentiate it from the traditional DSMC:
\begin{itemize}
\item[1.] As we already highlighted earlier, because of the complete isotropy of the system, there is no need to trace the directions of the particles' momenta. Initially, we start with randomly distributed directions, and the interactions preserve this randomness. As a conclusion, explicitly tracing the momenta would only add Monte-Carlo noise to the simulation.
\item[2.] The cell's EM plasma particles are collectively characterized by the temperature $T_{\text{cell}}$, while the neutrinos are represented by distinct simulation particles. This introduces a necessity for different treatment of these species. 
\item[3.] The interaction must obey quantum statistics. In particular, each interaction between neutrinos and EM plasma, resulting in the final states $(i,j)$ with the energies $E_{3},E_{4}$, is damped with the Pauli blocking factors 
\begin{equation}
    P_{\text{block}} = (1-f_{i}(E_{3}))(1-f_{j}(E_{4})) < 1,
    \label{eq:acceptance-pauli}
\end{equation}
where $f$ is the corresponding energy distribution inside the cell. In practice, it means that on top of the traditional DSMC pair acceptance criterion~\eqref{eq:vanilla-DSMC-acceptance}, there is the final acceptance factor $F$ post-simulating the collision kinematics.
\end{itemize}
On top of this, specific properties of weak interaction governing $\nu$-$\nu$ and $\nu$-EM interactions, such as scaling of the interaction cross-section $\sigma v \sim G_{F}^{2}s$ with the invariant mass of the interacting particles, require adjusting the pair's sampling compared to the pair sampling in the traditional DSMC.

Below, we summarize how these features are incorporated in the NTC kernel (see also Fig.~\ref{fig:new-NTC}).

\subsubsection{Treatment of EM particles inside the cell}

At the beginning of the simulation, the cell's EM temperature $T_{\text{cell}}$ matches the global temperature $T$ and gives the total initial energy of electrons and positrons available for the interaction with neutrinos, $\rho_{e^{\pm}}(T_{\text{cell}})$. It is calculated using the Fermi-Dirac distribution $f_{\text{FD}}(T,E)$, where $E = \sqrt{k^{2}+m_{e,\text{th}}^{2}}$. The thermal electron mass $m_{e,\text{th}}(T)$ has the form
\begin{equation}
    m_{e,\text{th}}^{2}(T) = m_{e}^{2}+\delta m^{2}_{e(2)}(T),
\end{equation}
with $\delta m^{2}_{e(2)}$ given by Eq.~\eqref{eq:thermal-electron-mass}.

The energies $E_{e}$ of the electrons and positrons may be retrieved from the same distribution $f_{\text{FD}}(T,E)$. Any interaction involving the EM sector (resulting in a change of the cell's EM energy $E_{\text{EM,cell}}$) updates the temperature according to the relation $T_{\text{cell}} = T(E_{\text{EM,cell}}/V_{\text{cell}})$, where the relation $T(\rho_{\text{EM}})$ is obtained by inverting Eq.~\eqref{eq:thermodynamic-relation}.

\subsubsection{Sampling pairs to interact}
Let us now specify details of the particles' sampling in $\nu$DSMC. The processes of the interactions between the neutrinos and the EM plasma are
\begin{align}
\nu_{\alpha}\bar{\nu}_{\alpha}&\to \nu_{\gamma}\bar{\nu}_{\gamma}, \quad \nu_{\alpha}\nu_{\beta}\to \nu_{\alpha}\nu_{\beta}, \quad \nu_{\alpha}\bar{\nu}_{\beta}\to \nu_{\alpha}\bar{\nu}_{\beta},  \\ 
\nu_{\alpha}e^{-}&\to \nu_{\alpha}e^{-}, \quad \nu_{\alpha}e^{+}\to \nu_{\alpha}e^{+}, \quad e^{+}e^{-}\to \nu_{\alpha}\bar{\nu}_{\alpha}
\label{eq:processes}
\end{align}
as well as their charge-conjugated versions. Here, $\alpha,\beta,\gamma = e,\mu,\tau$ with $\alpha \neq \gamma$. Given that in $\Lambda$CDM the lepton flavor asymmetries may be neglected, we do not introduce antineutrinos explicitly. Instead, when selecting a particular neutrino, we randomly assign it to a neutrino or an antineutrino of the given flavor.

Given the distinct properties of the EM particles and neutrinos, we find it convenient to split the interactions into three categories $I = 1,2,3$ and particle types $(i,j)$: 
\begin{enumerate}
\item Processes $\nu_{\alpha}\nu_{\beta}\to X$, 
\item Processes $\nu e \to \nu e$,
\item Processes $e^{+}e^{-}\to \nu\bar{\nu}$.
\end{enumerate}
For categories 1 and 2, $\nu$ stands for neutrino or anti-neutrino; $X$ denotes any possible final state. The total number of pairs to be sampled is $N_{\text{pairs}} = \sum_{I}N_{\text{pairs}}^{I}$, with $N_{\text{pairs}}^{I}$ being the number of pairs for each category:
\begin{equation}
    N_{\text{pairs}}^{(I)} = N_{I}\frac{(N_{I}-1)}{2} (\sigma v)^{(I)}_{\text{cell,max}}\frac{\Delta t}{V_{\text{cell}}}
    \label{eq:Npairs-category}
\end{equation}
Here, $(\sigma v)^{(I)}_{\text{cell,max}}$ is the maximal value of the total pair interaction cross-section in the cell, for the particles $k,k'\in I$. Explicitly,
\begin{equation}
    (\sigma v)^{(I)}_{\text{cell,max}} = \max_{k,k'\in I}\left(\sum_{X}(\sigma v)_{k,k' \to X}\right)
    \label{eq:upper-bound-cross-section}
\end{equation}
On each step $\overline{1,N_{\text{pairs}}}$, the quantities~\eqref{eq:Npairs-category} serve as weights to select the particles from the given category. Once the category is selected, the pairs $k,k'\in I$ are selected from the corresponding pool, and the final states of the scattering $k+k'\to X$ are generated proportionally to the corresponding cross-sections $(\sigma v)_{k,k' \to X}$. After any accepted interaction, the weights~\eqref{eq:Npairs-category} are recalculated. 

A generic feature of all the relevant processes is that their cross-sections have a universal upper bound in terms of the energies of the incoming particles $E_{1},E_{2}$, and the angle $\alpha$ between their 3-momenta (see Appendix~\ref{app:interactions} for details):
\begin{equation}
    (\sigma v)_{k+k'\to X} \leq \overline{(\sigma v)}_{k+k'\to X} = C_{ij,X} \cdot E_{1}E_{2}\cdot f(\alpha), 
    \label{eq:cross-section-individual-upper}
\end{equation}
Here, $i,j$ are particle species of $k,k'$; $f(\alpha) = (1-\cos(\alpha))^{2}$ for the category 1, and $f(\alpha) = (1-\cos(\pi))^{2} = 4$ for the categories 2 and 3. $C^{(I)}_{ij,X}$ is the constant depending on the species $i,j$ and the state $X$. These asymptotics typically correspond to the limit $\sqrt{s}/m_{e}\gg 1$. Introducing the maximal energy of the particle of the type $j$ inside the cell, $E^{(I)}_{i,\text{max}} = \max_{k \in i}(E_{k})$, the universal upper bound~\eqref{eq:upper-bound-cross-section} becomes
\begin{equation}
    (\sigma v)^{(I)}_{\text{cell,max}} = C^{(I)}_{\text{max}}\times E^{(I)}_{i,\text{max}}\cdot E^{(I)}_{j,\text{max}}\times 4,
    \label{eq:universal-upper-bound}
\end{equation}
where 
\begin{equation}
   C_{\text{max}}^{(I)} =  \max_{i,j}\left(\sum_{X}C^{(I)}_{ij,X}\right)
\end{equation}
The pair of particles is then selected uniformly. Namely, neutrinos' flavor and energy are drawn from the pool of simulation particles, while the electrons' and positrons' energy is sampled from the distribution $p^{2}f_{\text{FD}}(p,m_{e,\text{th}}(T))$. Finally, $\cos(\alpha)$ is uniformly sampled within the range $(-1,1)$. In the limit $E_{k},E_{k'}\gg m_{e}$, the pair's interaction acceptance criterion~\eqref{eq:vanilla-DSMC-acceptance} becomes
\begin{equation}
    P_{\text{acc}} = \frac{\sum_{X}C^{(I)}_{ij,X}}{C_{\text{max}}^{(I)}}\cdot \frac{E_{k}E_{k'}}{E^{(I)}_{i,\text{max}}E^{(I)}_{j,\text{max}}}\cdot\frac{(1-\cos(\alpha))^{2}}{4}
    \label{eq:acceptance-brute-force}
\end{equation}
However, using such a brute-force sampling is very inefficient. The issue is that $E^{I}_{i,\text{max}}$ is much larger than the typical energies $\langle E_{i}\rangle$. This is especially true in non-standard scenarios involving high-energy injected neutrinos. In addition, replacing $f(\cos(\alpha))$ with $4$ for the $\nu\nu\to X$ processes is too excessive, resulting in the additional suppression $(1-\cos(\alpha))^{2}/4$. As a result, the acceptance probability~\eqref{eq:acceptance-brute-force} is $\ll 1$, which results in a huge amount of excessive pair sampling and heavily influences the performance.

Instead, we consider the modified selection of neutrinos and $e^{\pm}$ particles, which allows us to replace the denominator of Eq.~\eqref{eq:acceptance-brute-force} with the upper bound Eq.~\eqref{eq:cross-section-individual-upper}. The functional dependence of the upper bound suggests drawing particles based on energy-weighted probability $\omega_{\text{sel},k} = E_{k}/E_{\text{total}}$, where $E_{\text{total}}$ is the total energy of the species $i$ to which $k$ belongs. In particular, for the $e^{\pm}$ particles, this is equivalent to sample the energy from the probability distribution $E^{2}pf_{\text{FD}}(E,T)$, with the electron's mass $m_{e,\text{th}}^{2}(T)$. Additionally, for the category $I = 1$ events, we may sample $\cos(\alpha)$ using the inverse CDF from the PDF $\sim (1-\cos(\alpha))^{2}$, which reduces the factor 4 from~\eqref{eq:universal-upper-bound} to $4/3$. 

Then, the standard DSMC factor $N_{I}(N_{I}-1)/2 \cdot (\sigma v)_{\text{cell,max}}$ should be replaced with the following sum $k,k'\in (i,j)^{I}$ (with $k\neq k'$):
\begin{multline}
    \sum (\sigma v)_{k+k'} \leq C^{(I)}_{\text{max}}\cdot \sum_{k\neq k'}\frac{E_{k}E_{k'}}{2} \cdot f(\alpha) = \\ = C^{(I)}_{\text{max}}\cdot \frac{(S_{1}^{2} - \delta_{ij}S_{2})}{2} \cdot f(\alpha),
    \label{eq:cross-section-individual-upper-bound}
\end{multline}
where the first and second moments of the energy distributions are
\begin{equation}
    S_{1} = \sum_{k}E_{k}, \quad S_{2} = \sum_{k}E_{k}^{2},
\end{equation}
and $\delta_{ij} = 1$ if the particles are drawn from the same species' pool, and 0 otherwise.\footnote{For the electrons and positrons, these moments are calculated from the exact Fermi-Dirac distributions.}

As a result, the number of pairs to be sampled is significantly reduced. For the category 1 events ($\nu\nu\to X$), given that we do not distinguish neutrinos and antineutrinos, we draw two particles from the same pool, so $i= j$. The first neutrino is selected with the probability $E_{1}/(E_{\nu,\text{total}})$, while the second neutrino with $E_{2}/(E_{\nu,\text{total}}-E_{1})$. Explicitly,
\begin{equation}
    N_{\text{pairs}}^{(1)} = C_{\text{max}}^{(1)}\times S \times \frac{4}{3}\times \frac{\Delta t}{V_{\text{cell}}}
    \label{eq:Npairs-category-1-optimized}
\end{equation}
Here, $S = \frac{1}{2}\left((S^{\nu}_{1})^{2}-S^{\nu}_{2}\right)$. For the events of categories 2 and 3, we have
\begin{align}
    N^{(2)}_{\text{pairs}} &= C^{(2)}_{\text{max}}\times S_{1}^{\nu}S^{e}_{1}\times 4\times \frac{\Delta t}{V_{\text{cell}}}, \\ 
    N^{(3)}_{\text{pairs}} &= 2C^{(3)}_{\text{max}}\times S_{1}^{e^{+}}S_{1}^{e^{-}}\times 4\times \frac{\Delta t}{V_{\text{cell}}},
    \label{eq:Npairs-category-3-optimized}
\end{align}
where a factor of $2$ comes from combinatorics.

The resulting acceptance probabilities for the pairs are calculated using Eq.~\eqref{eq:cross-section-individual-upper}:
\begin{equation}
    P_{\text{acc}} = \frac{(\sigma v)_{k+k'}}{C_{\text{max}}^{(I)}\cdot E_{k}E_{k'}\cdot f(\alpha)}\leq 1
    \label{eq:interaction-acceptance-optimized}
\end{equation}

\subsubsection{Sampling final state and Pauli blocking}

Once the pair $(i,j)$ is accepted for the interaction, the final state $X$ in the scattering $i,j\to X$ is selected based on the cross-sections $(\sigma v)_{ij\to X}(s,\cos(\alpha),m_{e}(T_{\text{cell}}))$. Here, $T_{\text{cell}}$ is the cell's EM temperature, while $m_{e}(T) = \sqrt{m_{e}^{2}+\delta m_{e(2)}^{2}}$ is the thermal electron mass, with the correction $\delta m_{e(2)}^{2}$ given by Eq.~\eqref{eq:thermal-electron-mass}.

The kinematics of the scattering 
\begin{equation}
k(E_{1})+k'(E_{2})\to X_{1}(E_{3})+X_{2}(E_{4}) 
\label{eq:scattering-explicit}
\end{equation}
is sampled in the following way. We start with the energies $E_{1},E_{2}$ and the $\alpha$ angle and restore the $s$ Mandelstam variable. To perform all computations in the center-of-mass reference frame, we express the energies of all particles in a Lorentz-invariant form via scalar products of their 4-momenta with the 4-velocity of the cosmic frame system. Then, we use the polar and azimuthal angles of the scattering in the CM frame, $\{\theta_{\text{CM}},\phi_{\text{CM}}\}$, to derive the energies $E_3$ and $E_4$ of the final states $\{X_{1}(E_{3}),X_{2}(E_{4})\}$ from the spherical cosine theorem. While $\phi_{\text{CM}}$ is isotropic, $\theta_{\text{CM}}$ depends on the process; we generate it in terms of $\cos(\theta_{\text{CM}})$ using the distribution $d\sigma/d\cos(\theta_{\text{CM}})$ and rejection sampling (see Appendix~\ref{app:interactions} for details).

Finally, after simulating the kinematics, the final state of the scattering~\eqref{eq:scattering-explicit} undergoes the final acceptance based on the Pauli blocking factor
\begin{equation}
 P_{\text{block}} =  (1-f_{X_{1}}(E_{3}))(1-f_{X_{2}}(E_{4})),
 \label{eq:pauli-blocking}
\end{equation}
where $f_{X}(E)$ is the local (in-cell) energy distribution function of the particle $X$.

Currently, for the seed of the neutrino distribution function in Eq.~\eqref{eq:pauli-blocking}, we use the Fermi-Dirac distribution with the effective temperature 
\begin{equation}
T_{\nu_{\alpha},\text{eff}} = \left(\frac{240\pi^{2}E_{\nu_{\alpha},\text{cell}}}{14V_{\text{cell}}}\right)^{\frac{1}{4}},
\end{equation}
where we included a factor of $2$ accounting for neutrinos and antineutrinos. For this reason, and also because of the collective description of the EM particles by the temperature $T$, we need a large number of neutrinos per cell, $N_{\nu,\text{cell}}\gg 1$. This distinguishes $\nu$DSMC from the traditional DSMC simulations, where $N_{\text{cell}}\simeq 10$~\cite{stefanov2019basic}.

This approximation is meaningful only if the effects from non-standard physics do not deviate the cell's neutrino distribution function significantly from the Fermi-Dirac shape, especially in the domain of low moments $p\lesssim 3T$. Once this assumption does not hold, it would be necessary to generalize the Pauli blocking factors to the case of a non-zero particle-antiparticle asymmetry and an arbitrary shape of the distribution. This can be achieved by determining the energy occupation levels at the start of the NTC routine and then recalculating them after each interaction.

To summarize, calculating pairs' acceptance and kinematics sampling are done with the help of simple matrix elements and integrated cross-section. The approach may be generalized to non-standard neutrino interactions without heavy work on the implementation of the phase space. This distinguishes $\nu$DSMC from the discretization approach, which utilizes analytic reduction of the dimensionality of the integration in the collision integral, which leads to a very complicated structure even in the $\Lambda$CDM case~\cite{Hannestad:1995rs}.

\subsubsection{Neutrino oscillations}

We implement the neutrino oscillations similar to how it is done in Refs.~\cite{Ruchayskiy:2012si,Sabti:2020yrt,Akita:2024nam}: averaging neutrino transitions over oscillation length and representing the probability to interact of a neutrino state $\alpha$ as a neutrino state $\beta$ by the stationary probability $P_{\alpha\beta}(T,E_{\nu})$. In $\nu$DSMC, we perform the neutrino oscillations once per cell, at the beginning of the NTC routine. The approach may be tightened by performing oscillations for every selected neutrino pair.

The oscillation probabilities are:
\begin{multline}
P_{\alpha\beta}^{\rm avg}(T,E_\nu)=\sum_{i=1}^3 \big|U^{\rm m}_{\alpha i}(T,E_\nu)\big|^2 \big|U^{\rm m}_{\beta i}(T,E_\nu)\big|^2\,,
       \label{eq:oscillation-probabilities}
    \end{multline}
where $U$ is the unitary Pontecorvo-Maki-Nakagawa-Sakata matrix,
    \begin{align}
        U = \begin{pmatrix}
        c_{12}c_{13} & s_{12}c_{13} & s_{13}\\
        -s_{12}c_{23}-c_{12}s_{13}s_{23} & c_{12}c_{23}-s_{12}s_{13}s_{23} & c_{13}s_{23}\\
        s_{12}s_{23}-c_{12}s_{13}c_{23} & -c_{12}s_{23}-s_{12}s_{13}c_{23} & c_{13}c_{23}
        \end{pmatrix}\, ,
        \nonumber
    \end{align}
with $c_{ij} = \cos(\theta^{\mathrm{m}}_{ij})$, $s_{ij} = \sin(\theta^{\mathrm{m}}_{ij})$. The thermally corrected mixing angles $\theta^{\mathrm{m}}_{ij}$ read~\cite{Strumia:2006db}
\begin{align}
      \theta^\mathrm{m}_{12} &= \frac{1}{2}\arctan\left[\frac{\sin(2\theta_{12})}{\cos(2\theta_{12})+\frac{A_\mathrm{MSW}}{\Delta m^2_\mathrm{sol}}}\right]\\
      \theta^\mathrm{m}_{13(23)} &= \frac{1}{2}\arctan\left[\frac{\sin(2\theta_{13(23)})}{\cos(2\theta_{13(23)})\pm\frac{A_\mathrm{MSW}}{\Delta m^2_\mathrm{atm}}}\right]\\
      A_\mathrm{MSW} &= \frac{16\zeta(3)\sqrt{2}G_{F}E_{\nu}^2T^4}{\pi m_{W}^2}\ ,
\end{align}
where $\theta_{ij}$ are the mixing angles in vacuum, $\pm$ corresponds to the normal or inverted neutrino hierarchy, $G_{F}$ is the Fermi coupling, and $m_{W}$ is the $W$ boson mass. 

The term $A_\mathrm{MSW}$ comes from neutrino self-energy, namely the charged current contribution to the $\nu_{e}$'s self-energy. Unlike the diagonal neutral-current corrections, it cannot be absorbed by the neutrino fields' redefinitions, and hence propagates to the oscillation probabilities. This correction dampens oscillation probabilities for high temperatures and neutrino energies, but for the thermal neutrino population, it gradually disappears at $T \lesssim 3\text{ MeV}$.

\section{Cross-checks and results}
\label{sec:results}

The resulting $\nu$DSMC approach has passed numerous tests. 

First, we have checked that in the absence of the expansion of the Universe, independently of the initial conditions, the system tends to the state characterized by thermal equilibrium. It is only possible if the equations of state of the plasma (Sec.~\ref{sec:thermal-corrections}), the number of sampled pairs~\eqref{eq:Npairs-category-1-optimized}-\eqref{eq:Npairs-category-3-optimized}, acceptance weights~\eqref{eq:interaction-acceptance-optimized}, and kinematics sampling are described fully consistently. Any mistake may either introduce unbounded energy draw to neutrinos or EM particles, or a systematic departure from the full thermal equilibrium. 

We characterize thermal equilibrium by the state when the distribution function of all neutrinos becomes $E^{2}f_{\text{FD}}(T_{\nu_{\alpha}},E)$, with the temperature $T_{\nu_{\alpha}}$ equal to the final EM plasma temperature. Additionally, every tiny systematic drift may be captured considering the quantity
\begin{equation}
    \delta \rho_{\nu}(t) \equiv \left(\frac{\rho_{\text{EM}}}{\rho_{\nu}}\right)_{\text{eq}}\frac{\rho_{\nu}}{\rho_{\text{EM}}}-1 = \frac{\rho_{\nu}}{\rho_{\nu,\text{eq}}}-1
    \label{eq:delta-rho}
\end{equation}
Here, $\left(\rho_{\text{EM}}/\rho_{\nu}\right)_{\text{eq}}$ has been obtained using Fermi-Dirac distribution for neutrinos, and $\mathcal{O}(e^{3})$ EM plasma energy density (recall Eqs.~\eqref{eq:EM-thermodynamics}). Under full equilibration and a sufficiently large number of neutrinos in cell $N_{\nu,\text{cell}}\gg 100$ (to minimize systematic effects), $\delta \rho_{\nu}$ fluctuates around $-0.02\%$. This is a defect of the current implementation, and it will be improved in the future. 

For the fixed number of simulation neutrinos $N$, the level of fluctuations around the central value determines the error bar of the predictions of the solver. For the number $N = 10^{7}$, the absolute value of the fluctuations in $\delta \rho_{\nu}$ is at the level of 0.05\%. 

Second, we have verified that the QED corrections obtained by our calculations match the actual results used in Refs.~\cite{Akita:2020szl,EscuderoAbenza:2020cmq}.

Third, we have performed the same tests of $\nu$DSMC against state-of-the-art approaches as in Ref.~\cite{Ovchynnikov:2024rfu} but including the effects of the electron mass. We compared our method with the integrated neutrino Boltzmann solver from~\cite{EscuderoAbenza:2020cmq,Cielo:2023bqp} and the unintegrated Boltzmann solver from Refs.~\cite{Akita:2020szl,Akita:2024ork}. Namely, we have considered the evolution of the setup with neutrinos having a thermal shape of the distribution but with temperatures $T_{\nu}>T$ at $T = 3\text{ MeV}$, and recovered results similar to Fig. 2 from~\cite{Ovchynnikov:2024rfu}. In addition, we studied the plasma equilibration by instantly injecting 70-MeV high-energy neutrinos with different flavor patterns into the primordial plasma at $T = 3\text{ MeV}$ and tracing their equilibration. The comparison shows very good agreement, both in terms of the integrated quantity $\delta \rho_{\nu}$ and the shape of the neutrino energy distribution.

\begin{figure}[t!]
    \centering
    \includegraphics[width=\linewidth]{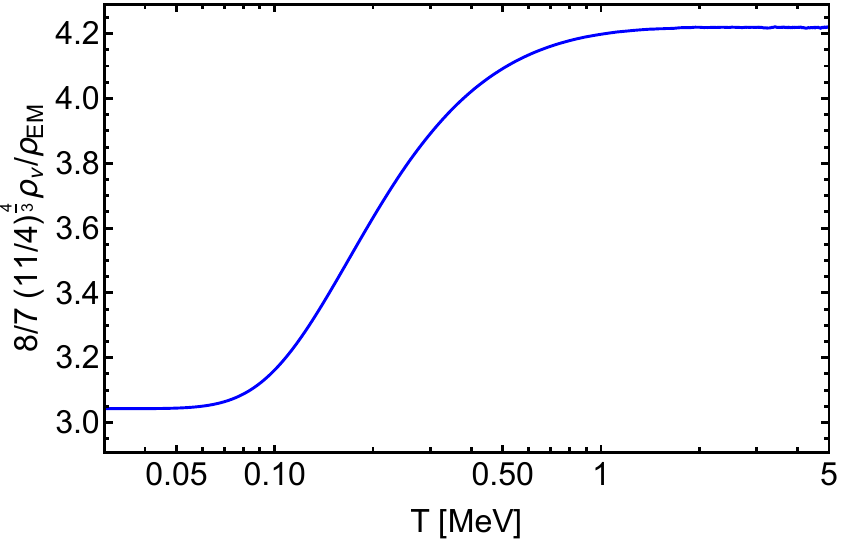}\\ \includegraphics[width=\linewidth]{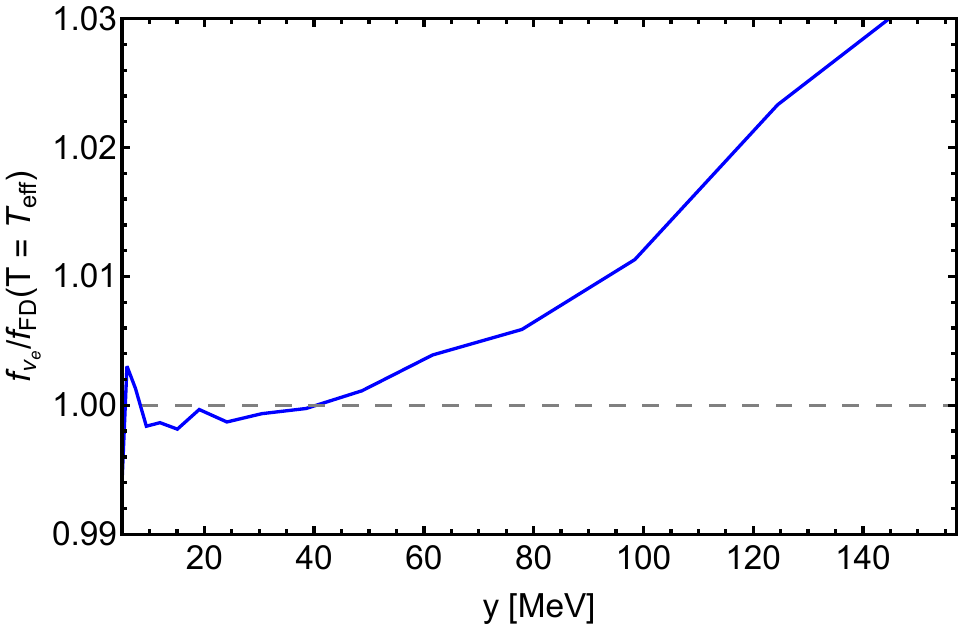}
    \caption{The evolution of the Universe in the temperature domain $T \in (30\text{ keV},10\text{ MeV})$ in $\Lambda$CDM as obtained within the $\nu$DSMC framework. Top panel: the temperature evolution of the ratio $8/7 \cdot (11/4)^{\frac{4}{3}}\rho_{\nu}/\rho_{\text{EM}}$, which tends to $N_{\text{eff}}$ in the limit of small temperatures $T\ll m_{e}$. Bottom panel: the ratio of the electron neutrino distribution function to the Fermi-Dirac distribution in terms of the comoving momentum $y = p \cdot (a(T)/a(T = 10\text{ MeV}))$. The Fermi-Dirac function is evaluated at temperatures $T_{\text{eff}}$ being the solution $E_{\nu_{e},\text{total}}/V_{\text{system}} = 3/2\zeta(3)/\pi^{2} T_{\text{eff}}^{3}$. The wiggles in the domain $y \lesssim 10$ are caused by limited statistics in this region (corresponding to $p/T \lesssim 0.02$).}
    \label{fig:results}
\end{figure}

Finally, to test the treatment of the expansion of the Universe inside $\nu$DSMC, Eqs.~\eqref{eq:expansion-discretized}, we have performed the DSMC simulation with the turned-off $\nu$-EM interactions, and calculated the final value of $N_{\text{eff}}$. Without the QED corrections included, we have obtained the expected result $N_{\text{eff}} = 3$. With the QED corrections, we have recovered $N_{\text{eff}} \approx 3.0104$, in agreement with Table~I of Ref.~\cite{EscuderoAbenza:2020cmq}.

\subsection{Results}

Let us now proceed to calculating $N_{\text{eff}}$. In $\nu$DSMC, it is defined as
\begin{equation}
    N_{\text{eff}} = \frac{8}{7}\left(\frac{11}{4}\right)^{\frac{4}{3}}\frac{E_{\nu,\text{tot}}}{E_{\text{EM,tot}}}\bigg|_{T = T_{\text{fin}}},
\end{equation}
where $E_{i,\text{tot}}$ is the total energy of the species $i$.
We start the evolution of the system at $T_{\text{ini}} = 10\text{ MeV}$, where neutrinos are in complete equilibrium with the EM particles, and evolve it down to $T_{\text{fin}} = 20\text{ keV}$, where electrons and positrons have already completely annihilated. We consider the total number of neutrinos $N = 3\cdot 10^{7}$, and the number of neutrinos per cell $N_{\nu,\text{cell}} = 4000$. For the timestep $\Delta t$, we considered
\begin{equation}
    \Delta t = \text{min}[0.01H^{-1}, \Gamma_{\text{max}}^{-1}],
\end{equation}
where $\Gamma_{\text{max}}$ is the estimate of the maximal possible interaction rate in the system. To eliminate the Monte-Carlo fluctuations (at the level of 0.03\% for the given $N$), we have launched the simulation several consecutive times. 

we have checked that changing $N_{\nu,\text{cell}}$ and $\Delta t$ only affect the results within tiny Monte-Carlo noise.

On top of that, we have utilized a few setups: with/without the QED corrections, and with/without neutrino oscillations. As a reference table of the expected values of $N_{\text{eff}}$, we use Tables~1 and~3 from Ref.~\cite{EscuderoAbenza:2020cmq}, together with Refs.~\cite{Gariazzo:2019gyi,Akita:2020szl,Froustey:2021azz,Drewes:2024wbw}. Collectively, these state-of-the-art studies report $N_{\text{eff}} = 3.034-3.035$ assuming the absence of the QED corrections and neutrino oscillations, and $N_{\text{eff}} = 3.043-3.044$ if the corrections and oscillations are included.

We obtain $N_{\text{eff}}=3.0345$ (no QED correction, no $\nu$ oscillations), $3.0427$ (QED on, oscillations off), and $3.0439$ (QED on, oscillations on). The error of our current implementation is within $\pm 0.02\%$, and it is obtained from the systematic drift of the quantity~\eqref{eq:delta-rho} in equilibrium. It may be further reduced if reducing the impact of $N_{\nu,\text{cell}}$-dependent effects and improving the implementation of Pauli principle~\eqref{eq:acceptance-pauli}. The results are in excellent agreement with the literature.

The temperature behavior of the dynamical analog of $N_{\text{eff}}$, $8/7\cdot(11/4)^{4/3}\rho_{\nu}/\rho_{\text{EM}}$, and the shape of the electron neutrino distribution function $f_{\nu_{e}}$ at $T_{\text{fin}}$ are shown in Figs.~\ref{fig:results}. $f_{\nu_{e}}$ is shown in terms of the variable $y = p\cdot a$. Its shape mostly aligns with the Fermi-Dirac distribution, being, however, slightly colder in the domain of small $p\ll T$ and hotter in the domain $p \gg T$, which agrees with the other studies~\cite{Gariazzo:2019gyi,Akita:2020szl}. This shape of the distortions corresponds to the non-thermal corrections due to the energy-dependent weak interaction rates~\cite{Ovchynnikov:2024rfu}; they are induced even in the absence of non-standard physics. Namely, the cross-section $\sigma v \propto s$, which means that high-energy neutrinos thermalize faster. When interacting, they drag neutrinos that populate typical energies. For the amount of neutrinos considered in our setup, the accuracy of DSMC allows smoothly tracing the neutrino distribution up to the energies $E>20T$.

As a final step, let us comment on the impact of our results. In $\Lambda$CDM, accurate calculations of $N_{\text{eff}}$ may be done even without solving the unintegrated Boltzmann equations -- approximating the neutrino distribution with the Fermi-Dirac one and reducing the task to evolving the neutrino temperature~\cite{Escudero:2018mvt,EscuderoAbenza:2020cmq,Cielo:2023bqp}. In addition, from the observational perspective, performing calculations of $N_{\text{eff}}$ with precision much better than $1\%$ may be overkill, given the claimed accuracy of current and upcoming CMB measurements and potential irreducible systematics that may pop up when interpreting measurements and spoil this accuracy. Despite this fact, the results are important for two reasons.

First, as a completely independent method to solve the neutrino Boltzmann equation, $\nu$DSMC is an independent cross-check of the existing approaches. Indeed, on the one hand, it does not impose any assumption about the shape of the neutrino distribution, unlike the integrated Boltzmann approach. On the other hand, unlike the state-of-the-art unintegrated methods, it does not rely on discretizing the comoving momentum space, simultaneously exactly conserving the energy of the system out of the box. Once the residual systematic shift in dynamic equilibrium has been further reduced, our method can be employed to study the impact of sub-leading effects on the plasma evolution. These include $\mathcal{O}(e^{4})$ corrections to the equation of state of the EM plasma~\cite{Escudero:2025kej} and NLO QED corrections to the weak scattering and annihilation rates, whose impact is investigated using various hard-thermal-loop resummation techniques and remains the subject of ongoing studies (see Refs.~\cite{Cielo:2023bqp,Jackson:2024gtr,Drewes:2024wbw} for details).

The second reason concerns the prospects for future applications of $\nu$DSMC. A key maturity test for any new method is its ability to reproduce established results. By demonstrating agreement with both integrated and unintegrated neutrino Boltzmann solvers for the evolution of neutrinos under non-thermal neutrino injections, as well as with the standard cosmological prediction for $N_{\text{eff}}$, the neutrino Monte Carlo method successfully passes this test. It opens up the possibility to robustly apply it in various non-standard scenarios that go beyond the simple setups.

\section{Conclusions}
\label{sec:conclusions}

Neutrino Direct Simulation Monte Carlo approach ($\nu$DSMC) is a new method to study the dynamics of the Early Universe at MeV temperatures in the presence of non-standard phenomena. Instead of explicitly solving the neutrino Boltzmann equation -- key ingredient and most complicated part in tracing the dynamics -- it simulates interactions of $N\gg 1$ neutrinos, electromagnetic (EM) particles, and non-standard species using Monte-Carlo techniques. 

$\nu$DSMC starts with a traditional DSMC used in the physics of rarefied gases (Sec.~\ref{sec:vanilla-DSMC}) and adds features of the MeV temperature Universe plasma, such as expansion of the Universe, homogeneity and isotropy, and instant thermalization of EM particles. The method is highly optimized, combining this with the possibility of using Monte-Carlo tools to generate the phase space of various reactions. Altogether, it makes it possible to apply $\nu$DSMC to a range of various scenarios, such as non-standard neutrino interactions and the presence of unstable species injecting non-thermal products into the plasma, where traditional methods to solve the Boltzmann equation experience complications.

In this paper, we have advanced the prototype of $\nu$DSMC proposed in~\cite{Ovchynnikov:2024rfu,Ovchynnikov:2024xyd} by consistently incorporating the finite electron mass and quantum QED corrections in Sec.~\ref{sec:nuDSMC}, and in particular in Subsecs.~\ref{sec:thermal-corrections} (QED corrections) and~\ref{sec:interaction-kernel} (interaction kernel). We have then applied the resulting framework to perform precision calculations of the effective number of relativistic degrees of freedom, $N_{\text{eff}}$, Sec.~\ref{sec:results}. Our result is $N_{\text{eff}} = 3.0439\pm 0.0006$, where the error comes from the Monte-Carlo noise of the setup we considered, and is in excellent agreement with the existing state-of-the-art approaches.  

The analysis we have performed is important for two reasons (Sec.~\ref{sec:results}). First, as a completely independent method, $\nu$DSMC cross-checks the state-of-the-art approaches of precision $N_{\text{eff}}$ calculations. Second, accurately reproducing the existing approaches for the setups where they are optimized with the Monte-Carlo method, we demonstrate that $\nu$DSMC is mature enough for applications in various non-standard scenarios, where the applicability of the state-of-the-art methods is limited. This is the subject of future work. 

\section*{Acknowledgements}

The authors thank Miguel Escudero for useful discussions about precision $N_{\text{eff}}$ calculations. MO received support from the European Union's Horizon Europe research and innovation programme under the Marie Sklodowska-Curie grant agreement No~101204216.

\bibliography{main.bib}

\newpage 
\appendix
\onecolumngrid

\section{Weak interactions of neutrino and EM particles}
\label{app:interactions}

In this section, we summarize the calculations of matrix elements and integrated cross-sections of the weak interactions between neutrinos and the EM particles.

We utilize the Fermi theory approximation of weak interactions, consider the processes~\eqref{eq:processes} with neutrinos and EM particles, and subsequently calculate squared matrix elements $\overline{|\mathcal{M}|^{2}}$ averaged over helicities and the total cross-section. We use the value of Weinberg's angle $\sin^{2}(\theta_{W}) = 0.23129$, Fermi coupling $G_{F} = 1.166378\cdot 10^{-11}\text{ MeV}^{-2}$, and electron's mass $m_{e} = 0.510998\text{ MeV}$~\cite{ParticleDataGroup:2024cfk}.

\begin{table}[h!]
    \centering
    \begin{tabular}{|c|c|c|c|c|c|}
    \hline Process & $\overline{|\mathcal{M}|^{2}}$ \\ \hline
 $\nu_e+\bar{\nu}_e\to e^{-}+e^{+}$ & $8 G_F^2 \left(m_e^2 \left(-g_A^2-4 g_A+g_V \left(g_V+4\right)\right) \left(p_{1}\cdot p_{2}\right)+\left(g_A-g_V\right)^2 \left(p_{1}\cdot p_{3}\right) \left(p_{2}\cdot p_{4}\right)+\left(g_A+g_V+4\right)^2
   \left(p_{1}\cdot p_{4}\right) \left(p_{2}\cdot p_{3}\right)\right)$ \\
 $\nu_e+e^{-}\to\nu_e+e^{-}$ & $4 G_F^2 \left(m_e^2 \left(g_A^2+4 g_A-g_V \left(g_V+4\right)\right) \left(p_{1}\cdot p_{3}\right)+\left(g_A-g_V\right)^2 \left(p_{1}\cdot p_{4}\right) \left(p_{2}\cdot p_{3}\right)+\left(g_A+g_V+4\right)^2 \left(p_{1}\cdot
   p_{2}\right) \left(p_{3}\cdot p_{4}\right)\right)$ \\
 $\nu_e+e^{+}\to\nu_e+e^{+}$ & $4 G_F^2 \left(m_e^2 \left(g_A^2+4 g_A-g_V \left(g_V+4\right)\right) \left(p_{1}\cdot p_{3}\right)+\left(g_A-g_V\right)^2 \left(p_{1}\cdot p_{2}\right) \left(p_{3}\cdot p_{4}\right)+\left(g_A+g_V+4\right)^2 \left(p_{1}\cdot
   p_{4}\right) \left(p_{2}\cdot p_{3}\right)\right)$ \\
 $\nu_{\alpha}+\bar{\nu}_{\alpha}\to\nu_{\alpha}+\bar{\nu}_{\alpha}$ & $128 G_F^2 \left(p_{1}\cdot p_{4}\right) \left(p_{2}\cdot p_{3}\right)$ \\
 $\nu_{\alpha}+\bar{\nu}_{\alpha}\to\nu_{\beta}+\bar{\nu}_{\beta}$ & $32 G_F^2 \left(p_{1}\cdot p_{4}\right) \left(p_{2}\cdot p_{3}\right)$ \\
 $\nu_{\alpha}+\bar{\nu}_{\beta}\to\nu_{\alpha}+\bar{\nu}_{\beta}$ & $32 G_F^2 \left(p_{1}\cdot p_{4}\right) \left(p_{2}\cdot p_{3}\right)$ \\
 $\nu_{\alpha}+\nu_{\alpha}\to\nu_{\alpha}+\nu_{\alpha}$ & $128 G_F^2 \left(p_{1}\cdot p_{2}\right) \left(p_{3}\cdot p_{4}\right)$ \\
 $\nu_{\alpha}+\nu_{\beta}\to\nu_{\alpha}+\nu_{\beta}$ & $32 G_F^2 \left(p_{1}\cdot p_{2}\right) \left(p_{3}\cdot p_{4}\right)$ \\
 $\nu_{\mu/\tau}+\bar{\nu}_{\mu/\tau}\to e^{-}+e^{+}$ & $8 G_F^2 \left(m_e^2 \left(g_V^2-g_A^2\right) \left(p_{1}\cdot p_{2}\right)+\left(g_A-g_V\right)^2 \left(p_{1}\cdot p_{3}\right) \left(p_{2}\cdot p_{4}\right)+\left(g_A+g_V\right)^2 \left(p_{1}\cdot
   p_{4}\right) \left(p_{2}\cdot p_{3}\right)\right)$ \\
 $\nu_{\mu/\tau}+e^{-}\to\nu_{\mu/\tau}+e^{-}$ & $4 G_F^2 \left(m_e^2 \left(g_A^2-g_V^2\right) \left(p_{1}\cdot p_{3}\right)+\left(g_A-g_V\right)^2 \left(p_{1}\cdot p_{4}\right) \left(p_{2}\cdot p_{3}\right)+\left(g_A+g_V\right)^2 \left(p_{1}\cdot
   p_{2}\right) \left(p_{3}\cdot p_{4}\right)\right)$ \\
 $\nu_{\mu/\tau}+e^{+}\to\nu_{\mu/\tau}+e^{+}$ & $4 G_F^2 \left(m_e^2 \left(g_A^2-g_V^2\right) \left(p_{1}\cdot p_{3}\right)+\left(g_A-g_V\right)^2 \left(p_{1}\cdot p_{2}\right) \left(p_{3}\cdot p_{4}\right)+\left(g_A+g_V\right)^2 \left(p_{1}\cdot
   p_{4}\right) \left(p_{2}\cdot p_{3}\right)\right)$ \\
 $\nu_{\mu/\tau}+e^{-}\to\nu_{\mu/\tau}+e^{-}$ & $4 G_F^2 \left(m_e^2 \left(g_A^2-g_V^2\right) \left(p_{1}\cdot p_{3}\right)+\left(g_A-g_V\right)^2 \left(p_{1}\cdot p_{4}\right) \left(p_{2}\cdot p_{3}\right)+\left(g_A+g_V\right)^2 \left(p_{1}\cdot
   p_{2}\right) \left(p_{3}\cdot p_{4}\right)\right)$ \\
 $e^{-}+e^{+}\to\nu_e+\bar{\nu}_e$ & $2 G_F^2 \left(\left(g_A+g_V+4\right) \left(m_e^2 \left(g_V-g_A\right) \left(p_{3}\cdot p_{4}\right)+\left(g_A+g_V+4\right) \left(p_{1}\cdot p_{3}\right) \left(p_{2}\cdot p_{4}\right)\right)+\left(g_A-g_V\right)^2
   \left(p_{1}\cdot p_{4}\right) \left(p_{2}\cdot p_{3}\right)\right)$ \\
 $e^{-}+e^{+}\to\nu_{\mu/\tau}+\bar{\nu}_{\mu/\tau}$ & $2 G_F^2 \left(\left(g_A+g_V\right) \left(m_e^2 \left(g_V-g_A\right) \left(p_{3}\cdot p_{4}\right)+\left(g_A+g_V\right) \left(p_{1}\cdot p_{3}\right) \left(p_{2}\cdot p_{4}\right)\right)+\left(g_A-g_V\right)^2
   \left(p_{1}\cdot p_{4}\right) \left(p_{2}\cdot p_{3}\right)\right)$ \\ \hline
    \end{tabular}
    \caption{Processes with neutrinos, electrons, and positrons we incorporate in $\nu$DSMC and their corresponding squared matrix elements averaged over the incoming particles' helicities. Here, $\alpha,\beta = e,\mu,\tau$, with $\alpha\neq \beta$. $g_{V} = -1+4\sin^{2}(\theta_{W})$, $g_{A} = -1$ are axial-vector and vector coupling in the neutral current of electrons. $p_{1}-p_{4}$ are 4-momenta in the process $1+2\to 3+4$.}
    \label{tab:placeholder}
\end{table}

The squared matrix elements for the processes in terms of the scalar products are listed in Table~\ref{tab:placeholder}. The expressions match with~\cite{Grohs:2015tfy}.

Expressing the scalar products in the center-of-mass (CM) frame, we may integrate them to obtain the cross-sections:
\begin{equation}
    \sigma =\frac{S}{8\pi v E_{1}E_{2}}\frac{|\mathbf{q}|}{\sqrt{s}}\int d\cos(\theta_{\text{CM}}) \overline{|\mathcal{M}|^{2}}, 
    \label{eq:cross-section}
\end{equation}
where $S$ is the symmetry factor accounting for the same particles in the final state ($S = 1$ is the particles are distinct, and $S=  1/2$ for the same particles), $E_{1},E_{2}$ are the energies of incoming particles, $v = \sqrt{(p_{1}\cdot p_{2})^{2}-m_{1}^{2}m_{2}^{2}}/E_{1}E_{2}$ is the relative velocity, $|\mathbf{q}|$ is the momentum of the outgoing particle in the CM frame, $s$ is the invariant mass of the collision, and $\theta_{\text{CM}}$ is the CM scattering angle.

The squared matrix elements expressed in terms of $s, m_{e}$, and $\cos(\theta_{\text{CM}})$ are used to sample the kinematics of the scattering in the CM frame inside the NTC kernel, recall Sec.~\ref{sec:interaction-kernel}.

\subsection{Upper bound on the cross-sections}
\label{sec:cross-sections-upper}
In the massless limit $s\gg m_{e}$, the cross-sections~\eqref{eq:cross-section} become
\begin{equation}
    (\sigma v)_{ab\to cd}\big|_{s\gg m_{e}} = \tilde{C}_{ab,cd}\cdot \frac{s_{ab}^{2}}{E_{a}E_{b}}
    \label{eq:asymptotics}
\end{equation}
In terms of $s$, these asymptotics are always larger than the corresponding finite-mass expressions, except for the $e^{+}e^{-}\to \nu_{e}\bar{\nu}_{e}$ case, for which the finite electron mass induces a positive $m_{e}^{2}/s$ correction. Fortunately, it may be compensated by multiplying the asymptotics $(\sigma v)_{e^{+}e^{-}\to \nu_{e}\bar{\nu}_{e}}\big|_{s_{ee}\gg m_{e}}$ by a constant factor $1.17$.

However, for pairs sampling, we need to find the upper bound not in terms of $s_{ab}$ but the energies $E_{a},E_{b}$ and $\cos(\alpha)$. In the massless limit, $s_{ab} = 2E_{a}E_{b}\cdot (1-\cos(\alpha))$, so Eq.~\eqref{eq:asymptotics} becomes
\begin{equation}
    (\sigma v)_{ab\to cd}\big|_{s\gg m_{e}} = \tilde{C}_{ab,cd}\cdot 2\cdot E_{a}E_{b}\cdot (1-\cos(\alpha))^{2}
    \label{eq:asymptotics-explicit}
\end{equation}
For $\nu\nu \to X$ scatterings, this is the universal upper bound of the finite-$m_{e}$ case, independently of $E_{1},E_{2},\alpha$. However, for the $\nu e \to \nu e$ and $e^{+}e^{-}\to \nu_{\alpha}\bar{\nu}_{\alpha}$ processes, the $(1-\cos(\alpha))^{2}$ scaling is spoiled with the $m_{e}/E$ terms, and for some values of $\alpha$, the asymptotics~\eqref{eq:asymptotics-explicit} may be significantly smaller than the exact cross-section~\eqref{eq:cross-section}. For these events, we replace $(1-\cos(\alpha))^{2}$ in Eq.~\eqref{eq:asymptotics-explicit} with its upper bound 4. 

Hence, the constants $C_{ij,X}$ and the functions $f(\alpha)$ entering Eq.~\eqref{eq:cross-section-individual-upper} are
\begin{equation}
   C_{ij,X} = \begin{cases}  2\cdot \tilde{C}_{ij,X}, \quad ij\to X\neq e^{+}e^{-}\to \nu_{e}\bar{\nu}_{e}, \\ 2\cdot 1.17\cdot \tilde{C}_{ij,X}, \quad ij\to X = e^{+}e^{-}\to \nu_{e}\bar{\nu}_{e},\end{cases}
\end{equation}
where $\tilde{C}_{ab,cd}$ are defined in Eq.~\eqref{eq:asymptotics}, and
\begin{equation}
    f(\alpha) = \begin{cases}
        (1-\cos(\alpha))^{2}, \quad \nu \nu \to X, \\ 4, \quad e\nu \to e \nu \ \text{ and } \ e^{+}e^{-}\to \nu\bar{\nu} 
    \end{cases}
\end{equation}

\end{document}